\documentclass[preprint,english,aps,pra,superscriptaddress]{revtex4}%
\usepackage{amsfonts}
\usepackage[T1]{fontenc}
\usepackage{amsmath}
\usepackage{amssymb}
\usepackage{babel}
\usepackage{graphicx}
\usepackage{subcaption}
\usepackage{xcolor}
\usepackage{lastpage}%

\begin{document}
	\title{Beyond the light-cone propagation of relativistic wavefunctions: numerical results}
 
	\author{X. Gutierrez de la Cal}
	\affiliation{Departmento de Qu\'imica-F\'isica, Universidad del Pa\' is Vasco, UPV/EHU,
		Leioa, Spain}
	\author{A.\ Matzkin}
	\affiliation{Laboratoire de Physique Th\'eorique et Mod\'elisation, CNRS Unit\'e 8089, CY
		Cergy Paris Universit\'e, 95302 Cergy-Pontoise cedex, France}

\begin{abstract}
It is known that relativistic wavefunctions formally propagate beyond the
light cone when the propagator is limited to the positive energy sector. By
construction, this is the case for solutions of the Salpeter (or relativistic
Schr\"odinger) equation or for Klein-Gordon \ and Dirac wavefunctions defined in
the Foldy-Wouthuysen representation. In this work we investigate
quantitatively the degree of non-causality for free propagation for different
types of wavepackets all having initially a compact spatial support. In the
studied examples we find that non-causality appears as a small transient
effect that can in most cases be neglected. We display several numerical
results and discuss the fundamental and practical consequences of our findings
concerning this peculiar dynamical feature.
\end{abstract}
\maketitle
\section{Introduction}

In classical relativistic physics, causality is associated with the light-cone
structure of Minkowski space-time: no event can be affected by an event lying
outside its past light-cone. In relativistic quantum mechanics, the situation
is more involved.\ It is indeed well-known that a relativistic evolution
driven by a positive energy Hamiltonian instantaneously turns an initial state
having compact spatial support into a distribution having mathematically a
non-zero amplitude everywhere in space \cite{hegerfeldt,kos,beck}.
Relativistic propagators restricted to the positive energy sector spill
outside the light cone \cite{greiner-qft}: it is only by including the
contribution of the anti-particle sector that a causal propagator is obtained.

While this observation points to the necessity of having antiparticles in a
relativistic quantum theory in order to preserve causality \cite{pad}, there
are instances in which no negative energies appear.\ For instance, in the
Salpeter equation \cite{salpeter-ref} (also known as the relativistic
Schr\"odinger equation), by construction the propagator is restricted to
positive energies. This also appears when the solutions of the Klein–Gordon or
Dirac equations are unitarily transformed in the Foldy–Wouthuysen (FW)
representation. Given the importance of the FW solutions—they are necessary
to obtain the classical limit, and it is sometimes claimed that densities
constructed from the FW wavefunctions are the only ones having a physical
meaning \cite{silenko}—it is instructive to investigate to which extent
there is an effective propagation outside the light cone. Indeed,
the instantaneous spreading of an initially localized wavefunction is a
mathematical fact, but it is often regarded as being physically irrelevant on
the ground that beyond the light-cone, propagation is extremely small and not
detectable in practice for any realistic physical state \cite{pavsic,ruijs}.

In this work, we will numerically investigate the fraction of the wavefunction
that effectively propagates outside the light-cone for different initial
states characterized by different parameters (width, mean momentum, and shape).
The common feature to the initial states we will employ is the requirement
that they have a compact spatial support. Up to now, most works that have
investigated this type of propagation, essentially in the context of the
Salpeter equation, have used states that have tails at $t=0$, such as initial
Bessel functions~\cite{usher} (one of the few cases for which analytical
solutions can be obtained) or Gaussian wavepackets
\cite{wiese,eckstein,pavsic,annalen}. If the states have initial compact
support, we can meaningfully and numerically follow the fraction that remains
inside the light cone as time evolves. We will see that although the fraction
spilling outside the light cone is small and does so during very short times,
it could have observational consequences for elementary~particles.

To this end, we will first (in Section \ref{sec2}) set the context by recalling in
which situations one is led to deal only with the propagator of the positive
energy sector. We will briefly recall in Section\ \ref{sec3} the arguments
proving the propagator is non-causal. We will give our results in Section
\ref{sec4}, describing the method employed and our choice of initial states.
We close the paper with a short discussion and conclusive comments (Section
\ref{sec-c}).

\section{Positive Energy Propagation\label{sec2}}

\subsection{Standard Relativistic Wave Equations}

The standard relativistic wave equations for spin-0 and spin-1/2 particles,
respectively, are the Klein–Gordon (KG) and Dirac equations,%
\begin{equation}
	i\hbar\partial_{t}\Phi=H\Phi\label{equu}%
\end{equation}
where $H$ represents the KG or Dirac Hamiltonians%
\begin{align}
	H_{KG}  &  =-\frac{\hbar^{2}}{2m}\left(  \sigma_{3}+i\sigma_{2}\right)
	\partial_{x}^{2}+mc^{2}\sigma_{3},\label{KG}\\
	H_{D}  &  =-i\hbar c\sigma_{1}\partial_{x}+mc^{2}\sigma_{3} \label{HD}%
\end{align}
and $\sigma_{i}$ are the usual Pauli matrices. We have used the Hamiltonian
form \cite{greiner} of the Klein–Gordon equation, for which $\Phi$ has two
components. We will be interested throughout this work in free propagation
along a single spatial direction; therefore, effectively restricting the
Hamiltonian to a spatial 1D problem: in this case, the Dirac spinor $\Phi$ has
only two nontrivial components, and this is why $H_{D}$ as given by Equation
(\ref{HD}) is two dimensional rather than four.

As is well-known \cite{greiner}, both $H_{KG}$ and $H_{D}$ admit positive and
negative energy solutions denoted $\Phi_{p}^{\pm}(t,x)$, with the $+$ and $-$
signs corresponding to positive and negative energy solutions. For instance,
for the Klein–Gordon equation we have \cite{wachter}
\begin{equation}
	\Phi_{p}^{\pm}(t,x)=\frac{1}{2\sqrt{mc^{2}E_{p}}}%
	\begin{pmatrix}
		mc^{2}\pm E_{p}\\
		mc^{2}\mp E_{p}%
	\end{pmatrix}
	e^{ipx/\hbar}e^{\mp iE_{p}t/\hbar}, \label{pwham}%
\end{equation}
where
\begin{equation}
	E_{p}=\sqrt{m^{2}c^{4}+p^{2}c^{2}}%
\end{equation}
and the prefactor is a normalization constant. The propagator $K(t,t^{\prime
};x,x^{\prime})$, evolving an initial state $\Phi(t^{\prime},x^{\prime})$ into
$\Phi(t,x),$ when expanded over the Hamiltonian eigenfunctions will contain
contributions from both the positive and negative states $\Phi_{p}^{\pm}%
(t,x)$. It is well-known~\cite{greiner-qft} that while the propagator is
causal, -- $K(t,t^{\prime};x,x^{\prime})$ vanishes for space-like separated
points, for which $x-x^{\prime}>c\left(  t-t^{\prime}\right)  $ -- ; the
restrictions $K^{\pm}(t,t^{\prime};x,x^{\prime})$ to an expansion over the
sole positive or negative energy eigenstates are not causal, in the sense that
$K^{\pm}(t,t^{\prime};x,x^{\prime})$ does not vanish for space-like separated
events. This is why it is often remarked that negative energies are necessary
in order to preserve relativistic causality \cite{pad}.

\subsection{The Salpeter or Relativistic Schr\"odinger Equation}

{The Salpeter equation \cite{salpeter}, also known as the relativistic
	Schr\"{o}dinger equation} or the Newton–Wigner–Foldy equation
\cite{horwitz,salpeter-ref}, describes a spinless particle obeying Equation
(\ref{equu}) with a Hamiltonian defined by%
\begin{equation}
	H_{S}=\sqrt{-\hbar^{2}c^{2}\partial_{x}^{2}+m^{2}c^{4}}. \label{HS}%
\end{equation}
Note that although this equation is obtained by formally quantizing the classical relativistic energy, the Salpeter equation was derived in \cite{salpeter} in a particular setting (see \cite{hist-salpeter} for more details). Due to the ambiguities of dealing with the differential inside the square root
operator, it is customary to work in momentum space since
\begin{equation}
	\sqrt{-\hbar^{2}c^{2}\partial_{x}^{2}+m^{2}c^{4}}\psi(t,x)=\frac{1}{\sqrt
		{2\pi\hbar}}\int dpE_{p}e^{ipx/\hbar}\psi(t,p).
\end{equation}
The plane-waves of positive energy $E_{p}$%
\begin{equation}
	\psi_{p}(t,x)=\exp\left(  -ipx/\hbar-iE_{p}t/\hbar\right)
\end{equation}
fulfill the relativistic Schr\"{o}dinger equation. By definition, $H_{S}$ is
positive definite so that the time evolution only includes a propagator
expanded over energy eigenstates, given by Equation (\ref{p1}) below.

Note that an arbitray initial wavefunction has a Fourier transform%
\begin{equation}
	\psi(0,x)=\frac{1}{\sqrt{2\pi\hbar}}\int dpe^{ipx/\hbar}\psi(0,p).
	\label{wdef}%
\end{equation}
By solving the evolution in momentum space, the time-evolved spatial
wavefunction is formally obtained as the Fourier transform%
\begin{equation}
	\psi(t,x)=\frac{1}{\sqrt{2\pi\hbar}}\int dpe^{ipx/\hbar}e^{-iE_{p}t/\hbar}%
	\psi(0,p). \label{TE}%
\end{equation}

\subsection{Foldy–Wouthuysen Density for the Klein–Gordon or Dirac Equation}

The solutions $\Phi_{p}^{\pm}(t,x)$ of the KG or Dirac equation in the
canonical representation (\ref{KG}) and (\ref{HD}) are well-known to give rise to
apparently curious properties (for example, the eigenvalues of the velocity are
always zero in the KG case and $c$ in the Dirac case \cite{greiner}; or, the
classical limit cannot be obtained as $\hbar\rightarrow0$ \cite{silenko}).
This is due to the fact that particle and anti-particle contributions
interfere even in the free case. The Foldy–Wouthuysen transformation \cite{FW,case}
is a unitary transformation in momentum space that separates particles from
anti-particles. For example, in the KG case the (pseudo-unitary) operator
\begin{equation}
	U=\frac{\left(  mc^{2}+E_{p}\right)  -\sigma_{1}\left(  mc^{2}-E_{p}\right)
	}{\sqrt{4mc^{2}E_{p}}} \label{FW-mat}%
\end{equation}
applied to the eigenstates $\Phi_{p}^{\pm}$ given by Equation (\ref{pwham}) lead
to
\begin{align}
	\Psi^{+}(t,p)  &  =U\Phi^{+}=\left(
	\begin{array}
		[c]{c}%
		1\\
		0
	\end{array}
	\right)  e^{-i\left(  E_{p}t-px\right)  /\hbar}\\
	\Psi^{-}(t,p)  &  =U\Phi^{-}=\left(
	\begin{array}
		[c]{c}%
		0\\
		1
	\end{array}
	\right)  e^{i\left(  E_{p}t-px\right)  /\hbar}%
\end{align}
and the transformed Hamiltonian is
\begin{equation}
	H_{FW}=UHU^{-1}=\sigma_{3}\sqrt{p^{2}c^{2}+m^{2}c^{4}}. \label{hamFW}%
\end{equation}
Similar relations for the Dirac solutions may be found in textbooks
\cite{greiner, wachter}.

The solutions $\Psi^{\pm}$ are indeed uncoupled: an initial particle state
$\Psi(0,x)=\left(  \psi(0,x),0\right)  $ has only an expansion over the
$\Psi^{+}(t,p)$ basis states, and thus only the upper component is non-zero.
Note that the $H_{FW}$ Hamiltonian is block diagonal, with each block consisting of
a Salpeter equation. We therefore see that if a density is defined from the
wavefunctions $\Psi\left(  t,x\right)  $ in the FW\ representation, then the
density of a positive energy state will be simply given by $\left\vert
\psi(t,x)\right\vert ^{2}$, which is precisely the density computed from the
Salpeter equation. We stress, however, that such a step involves defining a new
density that is different from the standard KG or Dirac densities (a unitary
transformation does not change the density nor the current). This new density
is free from the issues caused by the fact that the standard KG or Dirac
densities mix particles and anti-particles.\ For this reason, this density
displays several advantages and has been favored in some works
\cite{horwitz,silenko,pavsic,bohmian}, though it suffers from one important
drawback: it is formally non-causal.%1. These references 17 and 20 are not mentioned, please add.

\section{Non-Causality and Its Physical Implications\label{sec3}}

\subsection{Non-Causality of the Propagator}

The most straightforward way for showing the non-causality of the positive
energy propagator $K^{S}(t,t^{\prime};x,x^{\prime})$ associated with the
Salpeter Hamiltonian (\ref{HS}), or equivalently the positive component of Equation
(\ref{hamFW}), is to compute its expression. Indeed, by definition the
propagator should obey%
\begin{equation}
	\psi(t,x)=\int dx^{\prime}K^{S}(t,t^{\prime};x,x^{\prime})\psi(0,x^{\prime}).
	\label{prop}%
\end{equation}
Starting from Equation (\ref{TE}) and using the inverse transform of Equation
(\ref{wdef}), we immediately obtain
\begin{equation}
	K^{S}(t-t^{\prime};x-x^{\prime})=\frac{1}{2\pi\hbar}\int dpe^{ip\left(
		x-x^{\prime}\right)  /\hbar}e^{-iE_{p}\left(  t-t^{\prime}\right)  /\hbar}.
	\label{p1}%
\end{equation}
Different methods (see, e.g., \cite{wiese,redmount}) lead to the closed-form
expression

	\begin{equation}
		K^{S}(t-t^{\prime};x-x^{\prime})=\frac{im\left(  t-t^{\prime}\right)  }%
		{\pi\hbar\left(  \left(  x-x^{\prime}\right)  ^{2}-c^{2}\left(  t-t^{\prime
			}\right)  ^{2}\right)  ^{1/2}}K_{1}\left[  mc\left(  \left(  x-x^{\prime
		}\right)  ^{2}-c^{2}\left(  t-t^{\prime}\right)  ^{2}\right)  ^{1/2}%
		/\hbar\right]  \label{p2}%
	\end{equation}

where $K_{1}$ is a modified Bessel function of the second kind. From the
	asymptotic behavior $K_{1}(X)\sim\exp\left(  -X\right)  /\sqrt{X}$ for large
	$X$, it immediately follows that the propagator spills beyond the light cone.
	Recall that $K^{S}$, sometimes known as the Newton-Wigner propagator~\cite{nw}, is not Lorentz-invariant but does propagate the wavefunction, Equation
	(\ref{prop}), whereas a Lorentz-invariant propagator does not \cite{pad}. Note
	that the negative energy propagator displays the same behavior as $K^{S}$,
	with the value for space-like arguments being of opposite~sign.

Another line of reasoning relies on Paley-Wiener arguments. It can be proved
(see, e.g., \cite{hegerfeldt,kos,beck}) that for any semi-bounded Hamiltonian, a
wavefunction initially localized on a compact support immediately spreads
everywhere as soon as the evolution starts (or, conversely, any wavefunction
that remains bounded on a compact support must be zero everywhere).
Instructive illustrations of this theorem for simple wave equations were given
in Ref. \cite{karpov}.

\subsection{Physicality of Positive Energy Propagation}

The Salpeter equation, although it is attractive as it results from quantizing
the classical relativistic Hamiltonian, is usually regarded as an approximate
model for a spinless particle, correctly described by the Klein–Gordon
equation. Note, however, that for a neutral particle, the solutions of the
Klein–Gordon equation are real and can be combined to obtain a complex
wavefunction obeying the Salpeter equation \cite{pav1,pav2}.

The situation is more involved from the point of view of the FW
representation. The interpretational difficulties of the standard KG or Dirac
densities are related to the problems of defining a position operator. The
position operator $\hat{X}$ in the FW representation is equivalent to the
Newton-Wigner position operator \cite{nw} in the standard representation. On
this basis, it is often argued that the FW density is the physical one
\cite{silenko2}.\ The problem is then knowing how to cope with the non-causal nature
of the propagation.

The answer that has been given is, in a nutshell, that non-causality is in
practice undetectable. First, it has been argued that the probability of such
a detection is so low that it would be unlikely to detect such an event even
over billions of years \cite{ruijs}. Second, it is difficult to imagine how
signals could be sent superluminally since modulations cannot be produced from
the exponentially decaying tail \cite{pavsic}.\ It can also be remarked that
since non-causality is non-negligible only on distances of the order of a
Compton wavelength ($\lambda_{c}=\hbar/mc$) away from the light cone, a
detection on this scale is hardly feasible for elementary particles and
totally impossible for larger (not to mention macroscopic) bodies, for which
the Compton wavelength falls below the Planck length. The first step in
assessing whether these observations are plausible is to
quantitatively compute the fraction of the wavepacket that is propagated beyond the
light cone.

\begin{figure}[t]
	\begin{subfigure}[b]{0.49\textwidth}
		\includegraphics[width=\textwidth]{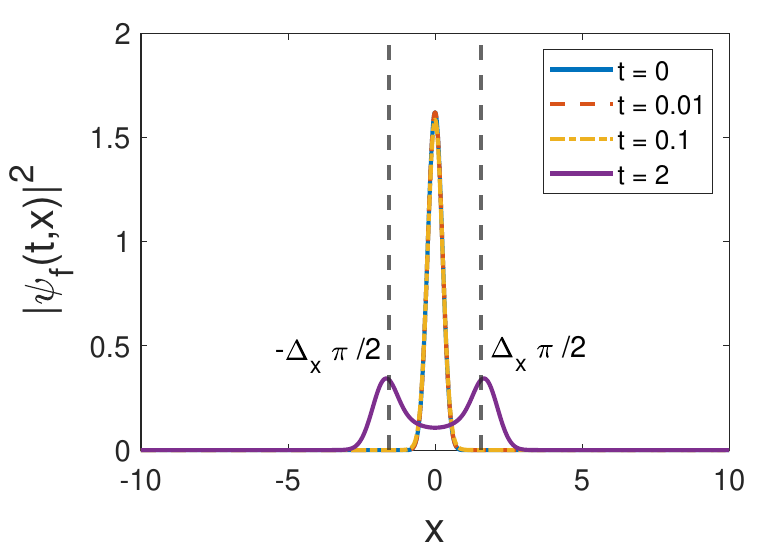}
		\centering
	\end{subfigure}
	\begin{subfigure}[b]{0.49\textwidth}
		\includegraphics[width=\textwidth]{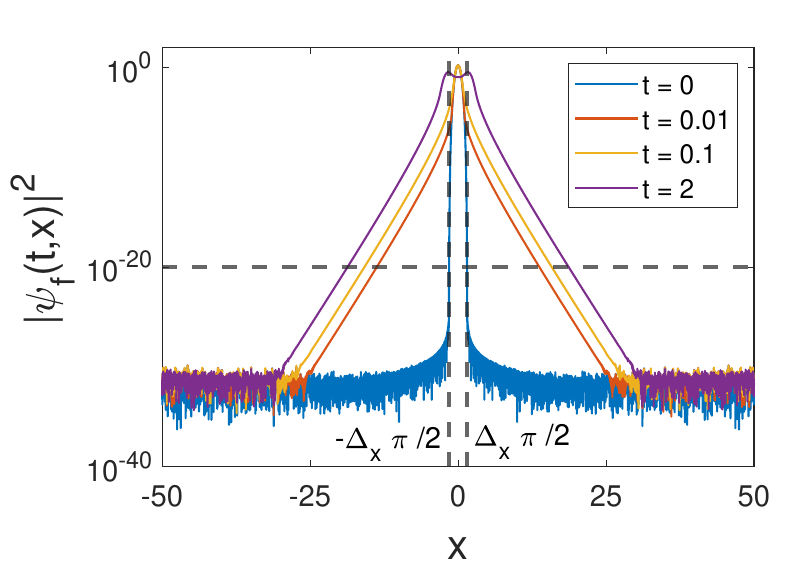}
		\centering
	\end{subfigure}
	\caption{$\left\vert \psi_{f}(t,x)\right\vert ^{2}$ for the function with
		compact support $f(x)=\cos^{8}(x/\Delta_{x})$ (with $\Delta_{x}=1$ and
		$p_{0}=0$) is shown for different values of $t$ (in natural units
		$\hbar=c=m=1$). \textbf{Left} panel displays the plots in the usual (linear) scale and
		the \textbf{right} panel shows the same quantities with a logarihtmic scale.\ The
		horizontal dashed line indicates the numerical zero. The initial wavefunction
		has compact support (no density above the numerical zero), whereas beyond the
		light-cone propagation appears for times $t>0$. }%
	\label{fig tails}%
\end{figure}

\section{Results\label{sec4}}

\subsection{Method}

The initial wave packet (WP) is defined by Equation (\ref{wdef}), which we rewrite
here as
\begin{equation}
	\psi(0,x)=\frac{1}{\sqrt{2\pi\hbar}}\int dpe^{ipx/\hbar}C(p;x_{0},p_{0})
\end{equation}
where $x_{0}$ and $p_{0}$ are the average position and momentum, respectively,
of our initial wave packet. We require $\psi(0,x)$ to have compact spatial
support.\ In this work, we will set $x_{0}=0$ and consider three different
initial WP of the form
\begin{equation}
	\psi_{f}(0,x)=(\theta(\pi\Delta_{x}/2-x)-\theta(x-\pi\Delta_{x}/2))e^{-ip_{0}%
		x}f(x) \label{initw}%
\end{equation}
with $f(x)$ given by $\cos^{8}(x/\Delta_{x})$, $\cos^{2}(x/\Delta_{x})$, or
$1$ (yielding a rectangular distribution). $\Delta_{x}$ gives the scale of the
spatial width of the packet, and $\theta(x)$ is the unit step function.
$\psi(0,x)$ is normalized to 1.

\begin{figure}[t]
	\begin{subfigure}[b]{0.49\textwidth}
		\includegraphics[width=0.9\textwidth]{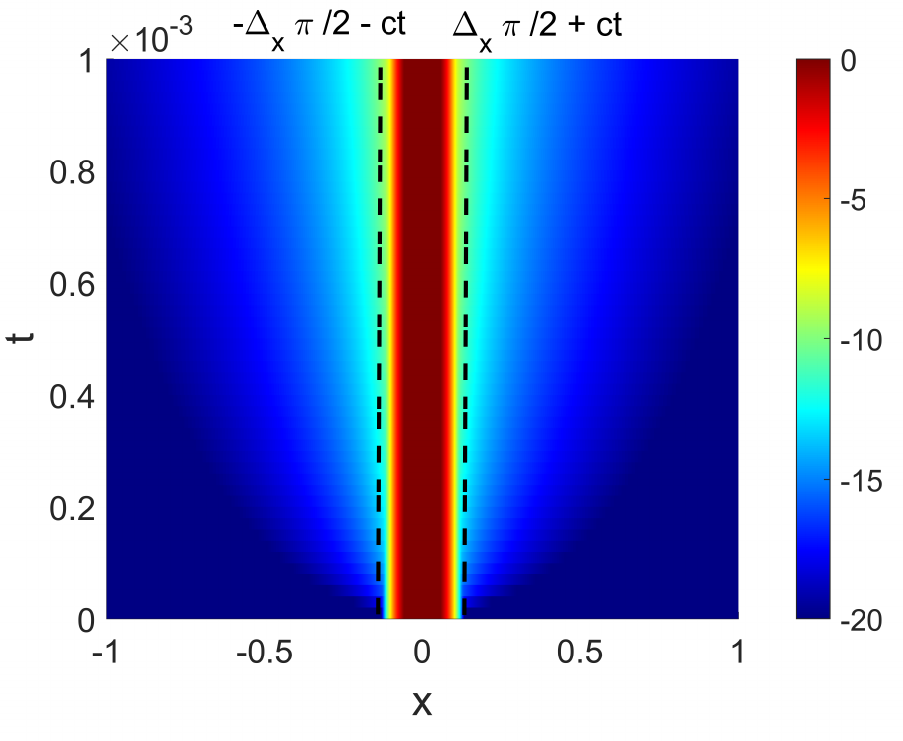}
		\centering
	\end{subfigure}
	\begin{subfigure}[b]{0.49\textwidth}
		\includegraphics[width=0.9\textwidth]{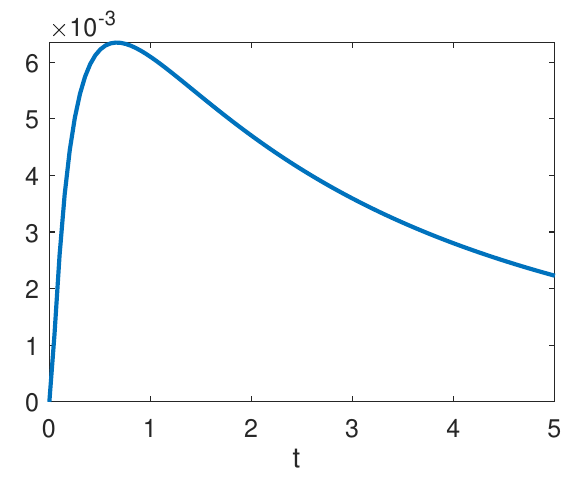}
		\centering
	\end{subfigure}
	
	\caption{\textbf{Left} panel: time evolution of $\log[\left\vert \psi_{f}
		(t,x)\right\vert ^{2}]$ for the function with initial compact support
		$f(x)=\cos^{8}(x/\Delta_{x})$ (with $\Delta_{x}=1$ and $p_{0}=0$) for short
		values of $t$. Note that the tails propagating beyond the light cone appear as
		soon as $t>0$. \textbf{Right} panel: Fraction of that same wave packet outside the
		light cone, where the transient aspect can be noticed.}%
	\label{fig tails2}%
\end{figure}

The initial momentum distribution is computed as the Fourier transform of the
compact wave function in coordinate space%

\begin{equation}
	C_{f}(p;x_{0},p_{0})=\frac{1}{\sqrt{2\pi\hbar}}\int_{-\Delta_{x}\pi/2}%
	^{\Delta_{x}\pi/2}\psi_{f}(0,x)e^{-ipx}dx
\end{equation}
hence, with the present notation, the time evolved wavefunction is given
by%
\begin{align}
	\psi_{f}(t,x)  &  =\int dx^{\prime}K^{S}(t,t^{\prime};x,x^{\prime})\psi
	_{f}(0,x^{\prime})\label{TEm}\\
	&  =\frac{1}{\sqrt{2\pi\hbar}}\int dpe^{ipx/\hbar}e^{-iE_{p}t/\hbar}%
	C_{f}(p;x_{0},p_{0}).
\end{align}

The initial momentum space wavefunctions $C_{f}(p;x_{0},p_{0})$ can be
obtained analytically,
\begin{equation}
	C_{cos^{m}}(p;x_{0},p_{0})=(2m!/\Delta_{x}^{m})\frac{\sin\left[  \left(
		p-p_{0}\right)  \Delta_{x}\pi/2\right]  \exp\left[  i\left(  p-p_{0}\right)
		x_{0}\right]  }{(p-p_{0})\prod_{n=1}^{n=m/2}(p-p_{0}-\frac{2n}{\Delta_{x}%
		})(p-p_{0}+\frac{2n}{\Delta_{x}})},
\end{equation}
when the initial state is a cosine function, and
\begin{equation}
	C_{1}(p;x_{0},p_{0})=\frac{2\sin[(p-p_{0})\Delta_{x}\pi/2]}{p-p_{0}%
	}e^{i\left(  p-p_{0}\right)  x_{0}}%
\end{equation}
for the case of the rectangular distribution for which $f(x)=1$. Note that the
simple poles appearing in the denominator are cancelled out by the sine in the numerator.

\begin{figure}[t]
	\begin{subfigure}[b]{0.49\textwidth}
		\includegraphics[width=0.9\textwidth]{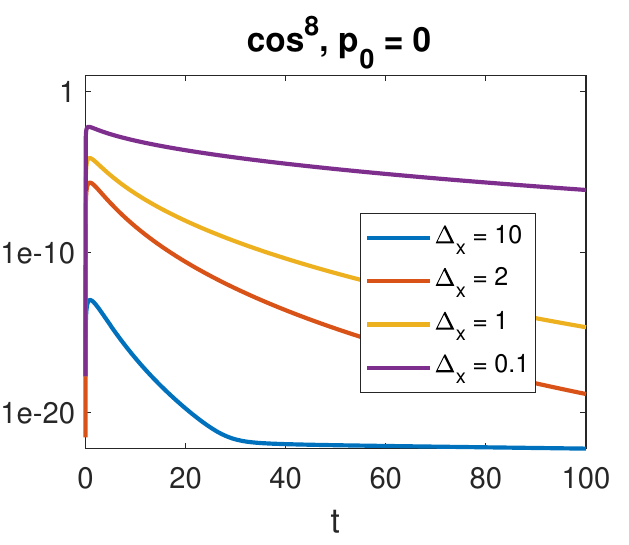}
		\centering
	\end{subfigure}
	\begin{subfigure}[b]{0.49\textwidth}
		\includegraphics[width=0.9\textwidth]{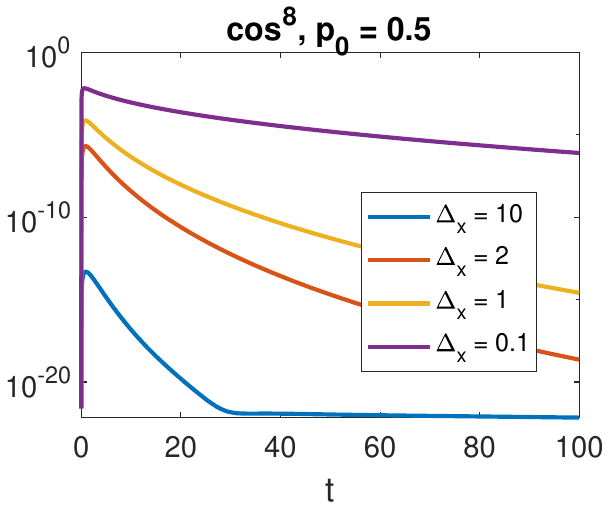}
		\centering
	\end{subfigure}
	\begin{subfigure}[b]{0.49\textwidth}
		\includegraphics[width=0.9\textwidth]{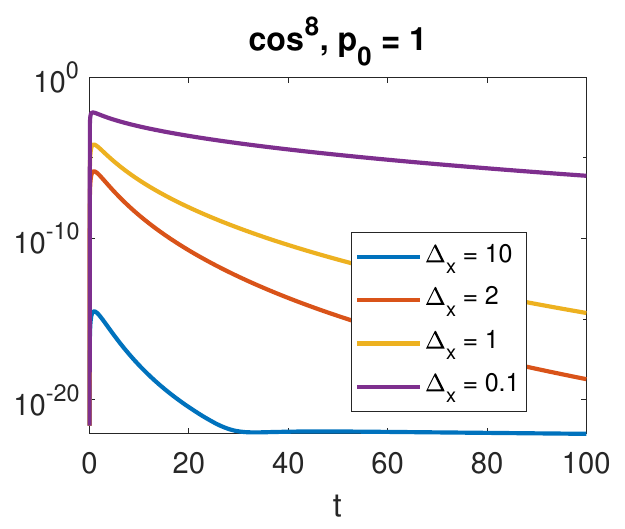}
		\centering
	\end{subfigure}
	\begin{subfigure}[b]{0.49\textwidth}
		\includegraphics[width=0.9\textwidth]{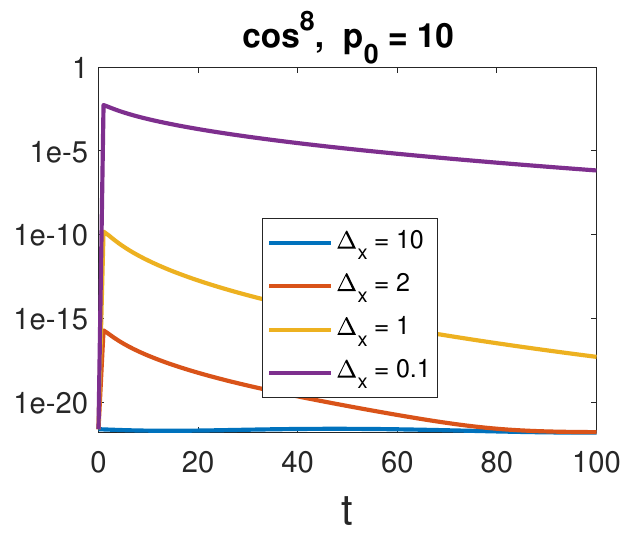}
		\centering
	\end{subfigure}
	
	\caption{ Fraction of the probability density lying outside the light cone for
		an initial wavepacket $\psi(0,x)$ [Equation (\ref{initw})] with $f(x)=\cos
		^{8}x/\Delta_{x}$, where $\Delta_{x}$ and $p_{0}$ vary as indicated.}%
	\label{fig cos8}%
\end{figure}

In practice, Equation (\ref{TEm}) is obtained numerically employing the trapezoidal
method in Matlab, with the bounds on the integration variable, $p_{i}$ and
$p_{f}$, taken large enough so that $C(p_{i,f};x_{0},p_{0})\sim0$.
Numerically, the actual computed value of the wave packet will not be exactly
zero outside its compact support (including at $t=0$), but a very small number
that should be smaller than our numerical zero, see Figure \ref{fig tails} for
the $f(x)=\cos^{8}(x/\Delta_{x})$ case for which the numerical zero is set at
$10^{-20}$. The calculations based on the method employed here have recently
been compared \cite{SR,PRA,AJP} to direct solutions of the relativistic wave
equations obtained by employing a high-precision finite-difference scheme,
resulting in an excellent agreement. In this section, all our results will be
given in natural units, $\hbar=c=m=1,$; hence, the Compton wavelength is
$\lambda_{C}=\hbar/mc=1$.

\begin{figure}[t]
	\begin{subfigure}[b]{0.49\textwidth}
		\includegraphics[width=0.9\textwidth]{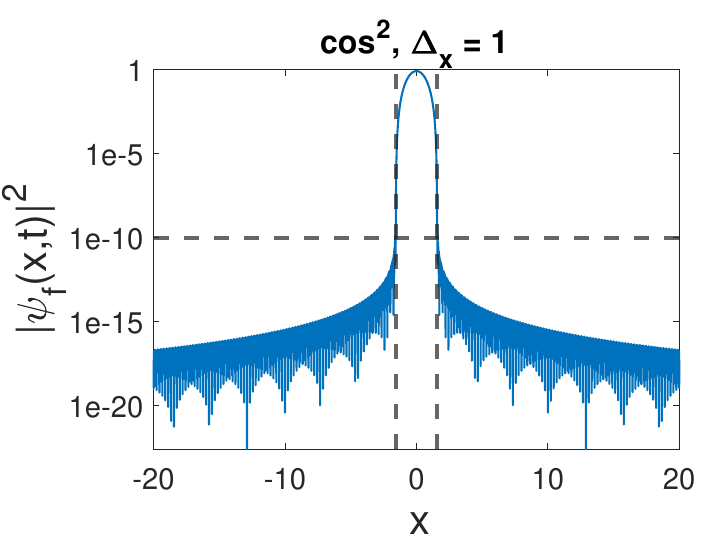}
		\centering
	\end{subfigure}
	\begin{subfigure}[b]{0.49\textwidth}
		\includegraphics[width=0.9\textwidth]{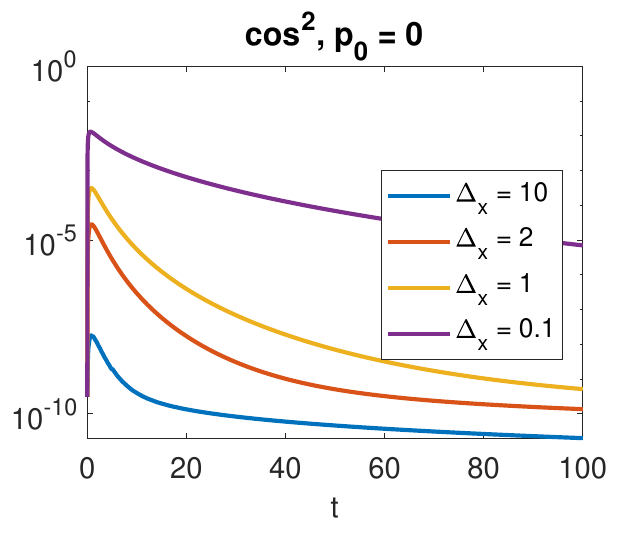}
		\centering
	\end{subfigure}
	
	\caption{\textbf{Left } panel: $\left\vert \psi_{f}(t=0,x)\right\vert ^{2}$ is plotted
		in logarithmic scale for the function with compact support $f(x)=\cos
		^{2}(x/\Delta_{x})$ (with $\Delta_{x}=1$ and $p_{0}=0$).\ \textbf{Right} panel:
		fraction of the probability density leaking outside the light cone as this
		initial wavepacket evolves (each curve shows the result for a different width
		$\Delta_{x}$).}%
	\label{fig cos2}%
\end{figure}

\subsection{Numerical Results}

The left panel of Figure \ref{fig tails} shows the probability density for a
typical initial wavefunction with compact support ($\psi_{f}(0,x)$ is given by
Equation (\ref{initw}) with $f(x)=\cos^{8}(x/\Delta_{x})$), along with a few
snapshots as the wavefunction evolves.\ On this scale, the tails propagating
outside the light cone are not visible, so we have plotted on the right panel
the same quantities on a logarithmic scale, clearly displaying beyond the
light-cone propagation (the light cone position is $\pm\left(  \pi/2+t\right)
$ in these units). Figure \ref{fig tails2} (left) shows a density plot for the
same initial wavefunction, while the right panel shows the fraction of the
density lying outside the light cone as time evolves. This fraction reaches a
maximum a very short time after initial propagation and then decreases to
zero for longer times. Figure \ref{fig cos8} shows the fraction of the
probability density outside the light cone for the same $\psi_{f}(0,x)$ but
with different initial widths $\Delta_{x}$ and initial momenta $p_{0}$.

\begin{figure}[t]
	\begin{subfigure}[b]{0.49\textwidth}
		\includegraphics[width=0.9\textwidth]{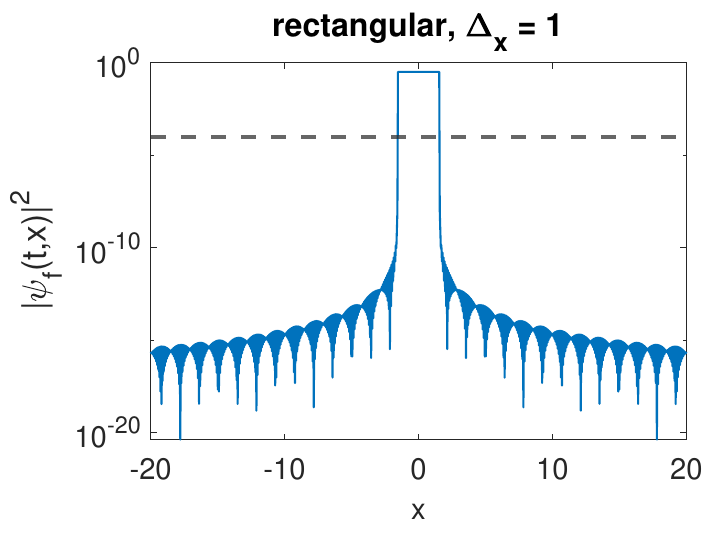}
		\centering
	\end{subfigure}
	\begin{subfigure}[b]{0.49\textwidth}
		\includegraphics[width=0.9\textwidth]{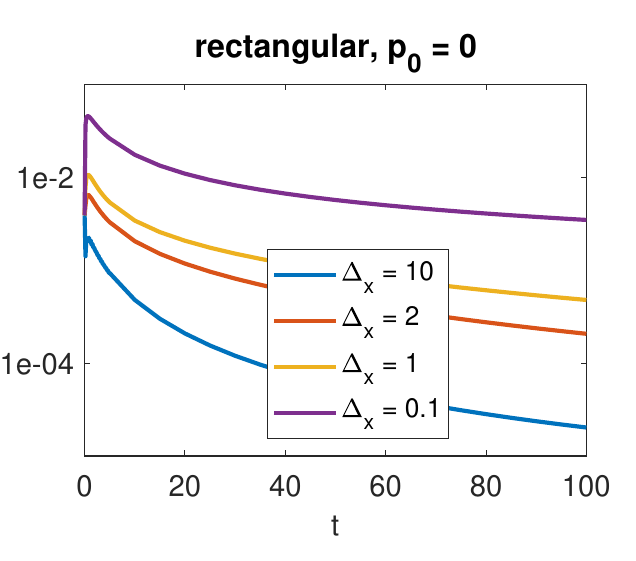}
		\centering
	\end{subfigure}%Please change the hyphen (-) into a minus sign (−, “U+2212”), e.g., “-1” should be “−1”. Reply: that is not possible, as we have already used minus signs, not hyphens, as generated by the software we use.
	\caption{Same as Figure \ref{fig cos2} but for an initial rectangle
		distribution, $f(x)=1$.}%
	\label{fig rect}%
\end{figure}

Figures\ \ref{fig cos2} and \ref{fig rect} display similar results but for an
initial wavefunction $\psi_{f}(0,x)$ with $f(x)=\cos^{2}(x/\Delta_{x})$ and
$f(x)=1$, respectively, (only the case with initial average momentum $p_{0}=0$
is shown). Note that we have taken a higher value for the numerical 0, given
that the momentum range over which we need to integrate in Equation (\ref{TEm}),
for each value of $t$, is significantly more extended than with a $\cos
^{8}(x/\Delta_{x})$ function. Finally, Figure \ref{fig compare} compares the
probability density propagating beyond the light cone for the three functions
$f(x)$ we have considered in this paper (with the same initial compact support
and mean momentum).

\begin{table}[b]
	\begin{center}%
		\begin{tabular}
			[c]{|c|c|c|c|c|}\hline
			& $\Delta_{x} = 10$ & $\Delta_{x} = 2$ & $\Delta_{x} = 1$ & $\Delta_{x} =
			0.1$\\\hline
			$cos^{8}$, $p_{0} = 0$ & - & - & - & $t \sim27 $\\\hline
			$cos^{2}$, $p_{0} = 0$ & - & - & $t \sim3.1$ & $t \sim43.5$\\\hline
			rectangular, $p_{0} = 0$ & $t \sim225$ & $t \sim1150$ & $t \sim2250$ & $t
			\sim9000 $\\\hline
		\end{tabular}
	\end{center}
	\caption{Time for which the fraction of the probability density propagating
		beyond the light cone remains above $10^{-4}$, for different initial
		wavefunctions.}%
	\label{table_p0}%
\end{table}

We have, in addition, included two tables. Table \ref{table_p0} specifies the
duration of significant superluminal propagation depending on $f(x)$ (the
compact support) and the width. We see that in the rectangular case for a
narrow wavepacket, a fraction ($0.01$\%) of the probability density remains
outside the light cone for times up to $t\approx10^{4}$ in natural units (for
an electron this corresponds to $1.3 \times10^{-17}$ s.). Table
\ref{table_pp0}, on the other hand, is interested in short times, reporting the
time at which the fraction of the probability density lying outside the light
cone is maximal, as well as the value of that fraction.

\begin{table}[ptb]
	\begin{center}%
		\begin{tabular}
			[c]{|c|c|c|c|c|}\hline
			wavefunction & $\Delta_{x} = 10$ & $\Delta_{x} = 2$ & $\Delta_{x} = 1$ &
			$\Delta_{x} = 0.1$\\\hline
			$cos^{8}$ & $t \sim0.96$ & $t \sim0.89$ & $t \sim0.84$ & $t \sim0.65$\\
			& $1.1\times10^{-13}$ & $2.15 \times10^{-6}$ & $7.25 \times10^{-5}$ & $1.38
			\times10^{-3}$\\\hline
			$cos^{2}$ & $t \sim0.84$ & $t \sim0.84$ & $t \sim0.80$ & $t \sim0.66$\\
			& $2.23 \times10^{-8}$ & $2.88 \times10^{-5}$ & $3.14 \times10^{-4}$ & $1.87
			\times10^{-3}$\\\hline
			rectangular & $t \sim0.64$ & $t \sim0.68$ & $t \sim0.66$ & $t \sim0.63$\\
			& $2.28 \times10^{-3}$ & $6.46\times10^{-3}$ & $1.06\times10^{-2}$ & $4.40
			\times10^{-2} $\\\hline
		\end{tabular}
	\end{center}
	\caption{Time for which the fraction of the probability density lying outside
		the light cone is maximal, along with the corresponding value of that
		fraction, for different initial wavefunctions (all with $p_{0} = 0$).}%
	\label{table_pp0}%
\end{table}

\section{Discussion and Conclusions\label{sec-c}}
As expected from mathematical arguments, our calculations confirm that the
	wavefunction propagates beyond the light cone as soon as $t>0$. Typical
wavepackets will have a spatial distribution $\Delta_{x}$ much larger than the
Compton wavelength $\lambda_{C}$.\ In this case, the propagation beyond the
light cone appears as a small transient effect, though not totally negligible
at short times (see the $\Delta_{x}=10$ curve in Figure \ref{fig rect}). Note
that short times, of the order of $t \approx1$ in the units used here, would
lie for an electron in the zeptosecond regime, a regime that is near
experimental reach \cite{zepto}. Of course for heavier particles, the time
scale scales inversely to the mass, so that for macroscopic bodies $t
\approx1$ would be shorter than the Planck time.

Even at considerably longer times, there is still a non-negligible probability
outside the light cone (see Table \ref{table_p0})). For narrower wavepackets,
the fraction of the probability distribution beyond the light cone can reach a
non-negligible percentage (see the $\Delta_{x}=0.1$ curve in Figures
\ref{fig cos8}--\ref{fig rect}). While it is generally
believed that the single particle formalism breaks down for wavepackets
narrower than $\lambda_{C}$ (but see \cite{bakke,hoffmann,su}), the propagating
wavepacket would still contribute to the one-particle sector of the
corresponding quantum-field theoretical description \cite{horwitz,pavsic}.
This does not necessarily entail that the superluminal propagation could
actually be detected, but, to the extent that the spatial density defined
above is the correct physical one, our results indicate this possibility must
be kept open.

\begin{figure}[b]
	\begin{subfigure}[b]{0.49\textwidth}
		\includegraphics[width=0.9\textwidth]{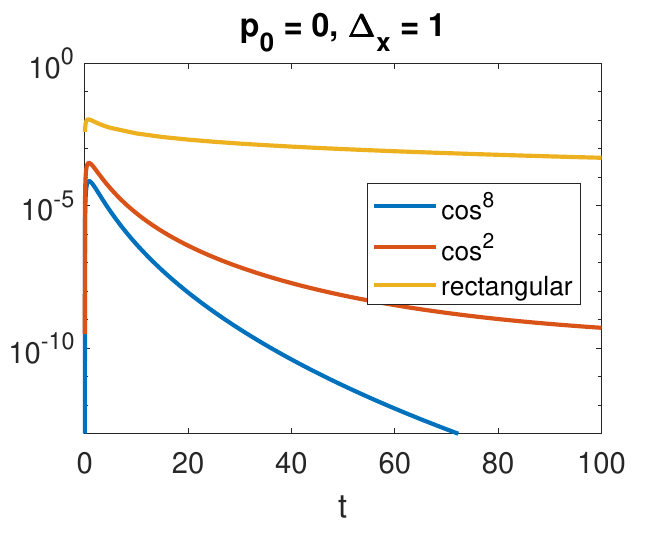}
		\centering
	\end{subfigure}
	\begin{subfigure}[b]{0.49\textwidth}
		\includegraphics[width=0.9\textwidth]{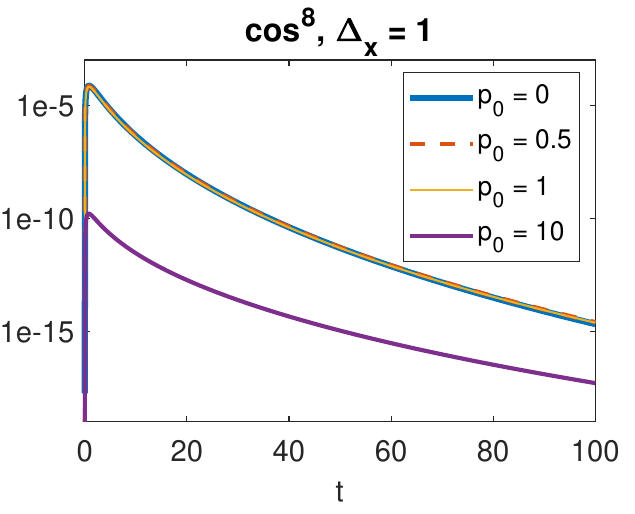}
		\centering
	\end{subfigure}%
	\caption{Fraction of the probability density leaking outside the light cone as
		time evolves. \textbf{Left} panel: Comparison of the three wave different initial
		compact supports $f(x)$ considered in this work, see Equation (\ref{initw}), with
		$p_{0}=0$ and $\Delta_{x}=1$. \textbf{Right} panel, for the case $f(x)=\cos
		^{8}(x/\Delta_{x})$, comparison of different values of the initial momentum
		$p_{0}$ (with $\Delta_{x}=1$ in all cases).}
	
	\label{fig compare}%
\end{figure}

Another interesting effect is the role of the initial average momentum. From
Figure~\ref{fig cos8}, one can see that the fraction of the probability density
leaking beyond the light-cone decreases as $p_{0}$ increases.\ Somewhat
paradoxically, wavepackets in the ultrarelativistic regime ($p_{0}\sim10$)
remain almost entirely within the light cone, while wavepackets with initial
zero momentum (whose average position does not move) are those that display
the highest proportion of superluminal propagation (see Figure \ref{fig compare}%
). It would be interesting to understand the reasons for this behavior. The
role of the shape of the wavepacket, which determines the range of the
contributing momenta in Equation (\ref{TEm}), is also interesting. From our choices
for $f(x)$, we can say that beyond the light cone, spreading increases as the
momentum range increases.\ Here too, there is no obvious argument to explain
this~behavior.

In summary, motivated by the issue of the physicality of the Foldy–Wouthuysen
(FW) density, we have numerically explored the time evolution of free
relativistic wavefunctions propagated by the sole positive energy propagator.
We have done so by choosing three different initial wavefunctions with compact
support and varying parameters. We conclude from our findings that beyond the
light cone, propagation is a very small effect that is non-negligible for
particles having a small mass (of the order of elementary particles) over a
short time after preparation of the initial state. It is, however, impossible to
assert that this effect cannot lead to any observational consequences if the
FW density is taken as being the physically meaningful quantity.


\vspace{1pt} 


	\begin{thebibliography}{99}                                                                                               %
		
		
		\bibitem {hegerfeldt}Hegerfeldt, G.C. Instantaneous spreading and Einstein
		causality in quantum theory. \emph{Ann. Phys. (Leipzig)} \textbf{1998}, \emph{7}, 716-725.%MDPI: Please add right pages. Reply OK
		
		\bibitem {kos}Kosinski, P. Salpeter Equation and Causality. \emph{Prog. Theor. Phys.}
		\textbf{2012}, \emph{128}, 59--65.
		
		\bibitem {beck}Beck, C. Local \emph{Quantum Measurement and Relativity};
		Springer Nature Switzerland: Cham, Switzerland, {2021}.
		
		\bibitem {greiner-qft}Greiner, W. \emph{Field Quantization}; Springer: Berlin, Germany,
		{1996}.
		
		\bibitem {pad}Padmanabhan, T. Obtaining the non-relativistic quantum mechanics
		from quantum field theory: Issues, folklores and facts. \emph{Eur. Phys. J. C}
		\textbf{2018}, \emph{78}, 563.
		
		\bibitem {salpeter-ref}Kowalski, K.; Rembielinski, J. Salpeter equation and
		probability current in the relativistic Hamiltonian quantum
		mechanics. \emph{\ Phys. Rev. A} \textbf{2011}, \emph{84}, 012108.
		
		\bibitem {silenko}Zou, L.; Zhang, P.; Silenko, A.J. Position and spin in
		relativistic quantum mechanics. \emph{Phys. Rev. A} \textbf{2020}, \emph{101}, 032117.
		
		\bibitem {pavsic}Pavsic, M. Localized States in Quantum Field Theory.
		\emph{Adv. Appl. Clifford Algebras} \textbf{2018}, \emph{28}, 89.
		
		\bibitem {ruijs}Ruijsenaars, S.N.M. On Newton-Wigner localization and
		superluminal propagation speeds. \emph{Ann.\ Phys.} \textbf{1981}, \emph{137}, 33--43.
		
		\bibitem {usher}Rosenstein, B.; Usher, M. Explicit illustration of causality
		violation: Noncausal relativistic wave-packet evolution. \emph{Phys.\ Rev.\ D}
		\textbf{1987}, \emph{36}, \ 2381.
		
		\bibitem {wiese}Al-Hashimi, M.H.; Wiese, U.J. Minimal position-velocity
		uncertainty wave packets in relativistic and non-relativistic quantum
		mechanics. \emph{Ann. Phys.} \textbf{2009}, \emph{324}, 2599--2621.
		
		\bibitem {eckstein}Eckstein, M.; Miller, T. Causal evolution of wave packets.
		\emph{Phys. Rev. A} \textbf{2017}, \emph{95}, 032106.
		
		\bibitem {annalen}Torre, A.; Lattanzi, A.; Levi, D. Time-Dependent
		Free-Particle Salpeter Equation: Numerical and Asymptotic Analysis in the
		Light of the Fundamental Solution. \emph{Ann. Der Phys.} \textbf{2017},
		\emph{529}, 1600231.
		
		\bibitem {greiner}Greiner, W. \emph{Relativistic Quantum Mechanics}; Springer:
		Berlin, Germany, {1996}.%MDPI: Please confirm if refs. 14 and 15 are same, if yes , please revise.
		% ANSWER they are NOT the same
		
		\bibitem {wachter}Wachter, A. \emph{Relativistic Quantum Mechanics}; Springer,
		Berlin, Germany,  {2011}.
		
		
		\bibitem {salpeter}Salpeter, E.E. Mass Corrections to the Fine Structure of
		Hydrogen-Like Atoms. \emph{Phys. Rev.} \textbf{1952}, \emph{87}, 328.
		
		\bibitem {horwitz}Rosenstein, B.; Horwitz, L.P. Probability current versus
		charge current of a relativistic particle. \emph{\ J. Phys. A Math. Gen.}
		\textbf{1985}, \emph{18}, 2115.
		
		
		
		\bibitem {hist-salpeter}Lucha, W.; Schoeberl, F.F. All Around the Spinless Salpeter Equation. \emph{arXiv} \textbf{1994}, arXiv:hep-ph/9410221.
		
		
		\bibitem {FW}Foldy, L.L.; Wouthuysen, S.A. On the Dirac Theory of Spin 1/2
		Particles and Its Non-Relativistic Limit. \emph{Phys. Rev.} \textbf{1950}, \emph{78}, 29.
		
		\bibitem {case} Case, K.M. Some Generalizations of the Foldy–Wouthuysen Transformation. \emph{Phys. Rev. } \textbf{1954}, \emph{95}, 1323.%1. These references 17 and 20 are not mentioned, please add. Reply: Corrected, thank you
		
		\bibitem {bohmian}Alkhateeb M.; Matzkin A. Relativistic Bohmian Trajectories
		and Klein–Gordon Currents for Spin-0 Particles. \emph{Found. Phys.}
		\textbf{2022}, \emph{52}, 104.
		
		\bibitem {redmount}Redmount, I.H.; Suen, W.M. Path integration in
		relativistic quantum mechanics. \textit{Int. J. Mod. Phys. A} \textbf{1993}, \emph{8}, 1629--1635.
		
		
		\bibitem {karpov}Karpov, E.; Ordonez, G.; Petrosky, T.; Prigogine, I.; Pronko, G.
		Causality, delocalization, and positivity of energy. \emph{Phys.\ Rev.\ A}
		\textbf{2000}, \emph{62}, 012103.
		
		\bibitem {pav1}Pavsic, M. Manifestly covariant canonical quantization of the scalar field and particle localization.
		\emph{Mod. Phys. Lett. A} \textbf{2018}, \emph{33}, 1850114.
		
		\bibitem {pav2}Pavsic, M. A new perspective on quantum field theory revealing possible existence of another kind of fermions forming dark matter. \emph{Int. J. Geom. Meth. Mod. Phys.} \textbf{2022}, \emph{19}, 2250184.
		
		
		
		
		\bibitem {nw}Newton, T.D.; Wigner, E.P. Localized States for Elementary
		Systems. \emph{Rev. Mod. Phys.} \textbf{1949}, \emph{21}, 400.
		
		\bibitem {silenko2}Silenko, A.J.; Zhang, P.; Zou, L Reply to Comment on
		\textquotedblleft Relativistic Quantum Dynamics of Twisted Electron Beams in
		Arbitrary Electric and Magnetic Fields. \emph{Phys. Rev.
			Lett.} \textbf{2019}, \emph{122}, 159302.
		
		\bibitem {SR}Guti\'{e}rrez de la Cal, X.; Alkhateeb, M.; Pons, M.;  Matzkin, A.; Sokolovski, D.  Klein
		paradox for bosons, wave packets and negative tunnelling times. \emph{Sci. Rep.}
		\textbf{2020}, \emph{10}, 19225.
		
		\bibitem {PRA}Alkhateeb, M.; Gutierrez de la Cal, X.; Pons, M.; Sokolovski, D.;
		Matzkin, A. Relativistic time-dependent quantum dynamics across supercritical
		barriers for Klein–Gordon and Dirac particles. \emph{Phys. Rev. A}
		\textbf{2021}, \emph{103}, 042203.
		
		\bibitem {AJP}Alkhateeb, M.; Matzkin, A. Relativistic spin-0 particle in a box:
		Bound states, wave packets, and the disappearance of the Klein paradox.
		\emph{Am.\ J.\ Phys. }\textbf{2022}, \emph{90}, 297.
		
		\bibitem {zepto}Mourou, G.; Mironov, S.; Khazanov, E.; Sergeev, A. Single cycle
		thin film compressor opening the door to Zeptosecond-Exawatt physics. \emph{
			Eur. Phys. J. Spec. Top. } \textbf{2014}, \emph{223}, 1181.
		
		
		
		
		\bibitem {bakke}Bakke, F.; Wergeland, H. Wave packets of relativistic electrons.
		\emph{Physica} \textbf{1973}, \emph{69}, 5--11.
		
		\bibitem {hoffmann}Hoffmann, S.E. The minimum width in relativistic quantum
		mechanics. \emph{J. Phys. B} \textbf{2018}, \emph{51}, 165302.
		
		\bibitem {su}Krekora, P.; Su, Q.; Grobe, R. Relativistic Electron
		Localization and the Lack of Zitterbewegung. \emph{Phys.\ Rev.\ Lett.}
		\textbf{2004}, \emph{93}, 043004
\end{thebibliography}
\end{document}